\documentclass[12pt,preprint]{aastex}

\begin{document}

\title{No Heavy Element Dispersion in the Globular Cluster M92
\altaffilmark{1}}

\author{Judith G. Cohen\altaffilmark{2}  }

\altaffiltext{1}{Based in part on observations obtained at the
W.M. Keck Observatory, which is operated jointly by the California
Institute of Technology, the University of California, and the
National Aeronautics and Space Administration.}

\altaffiltext{2}{Palomar Observatory, Mail Stop 249-17,
California Institute of Technology, Pasadena, Ca., 91125,
jlc@astro.caltech.edu}

\begin{abstract}

Although there have been recent claims that there is a large
dispersion in the abundances of the heavy neutron capture
elements in the old Galactic globular cluster M92, we show
that the measured dispersion  for the absolute abundances of four of the 
rare earth elements within
a sample of 12 luminous red giants in M92 ($\leq$ 0.07~dex)
does not exceed the
relevant sources of uncertainty.  As expected from previous studies,
the heavy elements show the  signature of the $r$-process.
Their abundance ratios
are essentially identical to those of M30, another nearby globular
cluster of similar metallicity.

\end{abstract}


\section{Introduction \label{section_intro} }

Globular clusters are canonically viewed as 
stellar systems within which all stars formed with the same
initial chemical inventory at approximately the same time
in the distant past.  More than 30 years ago it was realized
that the light elements (C, N, O, Na, Mg, and Al) show
correlated star-to-star variations \citep[see, e.g. the
review by][ and references therein]{gratton_araa}.  
There are a few massive globular clusters
which show a range of abundances
among both the light and heavy elements that  are believed to be remnants of
accreted satellites including 
$\omega$ Cen, whose anomalies were discovered more
40 years ago (see Norris, Freeman \& Mighell 1996 and
references therein),
and more recently M22 \citep{costa09,marino09}.
M15 was discovered by
\cite{m15_1997,sneden00} to show a range
of abundances among the heavy neutron capture elements,
confirmed by \cite{subaru_m15}.
In M15, a massive low-metallicity Galactic globular cluster, one
sees the $r$-process ratios among the heavy neutron capture elements,
with a wide range of $r$-element abundances while
the Fe-peak elements have constant abundances.
The decoupling of the heavy neutron capture elements from
the Fe-peak is common among extremely metal-poor halo
field stars, as was first shown by \cite{mcwilliam95},
see  e.g. Fig.~17 of \cite{cohen_umi} for a more recent view; it
commences at a metallicity comparable to 
that of M15. \cite{yong08} found variations
 of the $s$-process elements Zr and La in NGC~1851 which presumably
 have a different origin and are more easily explained in conventional
 globular cluster models.

M92, a Galactic globular cluster  with comparable
metallicity, distance, and age as M15, is slightly less luminous.
Until very recently, there was
no hint in the literature of any star-to-star variations for any element
heavier than Al in the many published  spectroscopic studies of
red giant
stars in M92
beginning with \cite{cohen79}.  \cite{sneden00}
carried out the most recent study which included
a substantial number of members of M92.
 
Very recently \cite{roe11} (RS11) claim to have detected
a large range ($\sim$0.6~dex for Eu) 
in the heavy neutron capture elements
in M92 with a sample of 19 stars observed with
the Hydra spectrograph at the WIYN 3.5~m telescope. 
Their choice of this instrument requires
balancing
the advantage of substantial multiplexing of objects against a restricted
spectral range and a moderate spectral resolution.  

In the present paper we present evidence that there is no
detectable range in the heavy neutron capture element
abundances for a sample of 12 giants in M92 with considerably
better spectra than those used by RS11.  Our measured
dispersion in abundance
for each of Y, Ba, La, and Eu, as well as for Fe, is ${\leq}$ 0.07~dex,
which as we will show
is comparable to that expected  from
known sources of uncertainty.  This is 
smaller by a factor of at least 4 than the
range claimed by RS11.

\section{Stellar Sample, Observations, Abundance Analyses}

To evaluate the validity of the
claim of star-to-star variation in the neutron capture element
abundances,  we examined the spectra of
luminous M92 giants with high S/N that were taken by us after the
HIRES detector upgrade in early 2004.   We had 7 such spectra;
five from April 2006 and two from June 2008.
Recently we observed 6 luminous RGB stars in M92
selected from
RS11 to  span a large range in their
claimed [Eu/Fe].  This group of stars includes 
one previously observed in April, 2006, for a total sample
of 12 luminous RGB stars in M92.  All  are members of M92
based on their radial velocities and the appearance of their spectra.
Details of our sample  are given in 
Table~\ref{table_m92_sample}.  

The spectra of these M92 stars are all of high S/N with
spectral resolution of 35,000. 
The April 2006 spectra were taken with HIRES-B; they
do not extend redder than 5950~\AA.  Since we took them to
get good coverage in the UV, their S/N at the wavelengths of
interest here is very high, exceeding 150 per spectral
resolution element at all relevant wavelengths.
The other M92 spectra were taken with HIRES-R.  They
cover 3900 to 8350~\AA.  
The 2011 spectra also have
very high S/N except for M92 IV-10, which has a somewhat lower
S/N.  The April 2008 spectra of the two faintest stars have
somewhat lower S/N, perhaps 80 per spectral resolution element.
There are 
small gaps between the three CCDs that are mosaiced together
to form the HIRES detector.  One key line of La~II, at 4086~\AA, 
falls into one of those gaps for the HIRES-B spectra only.

\begin{figure}
\epsscale{1.0}
\plotone{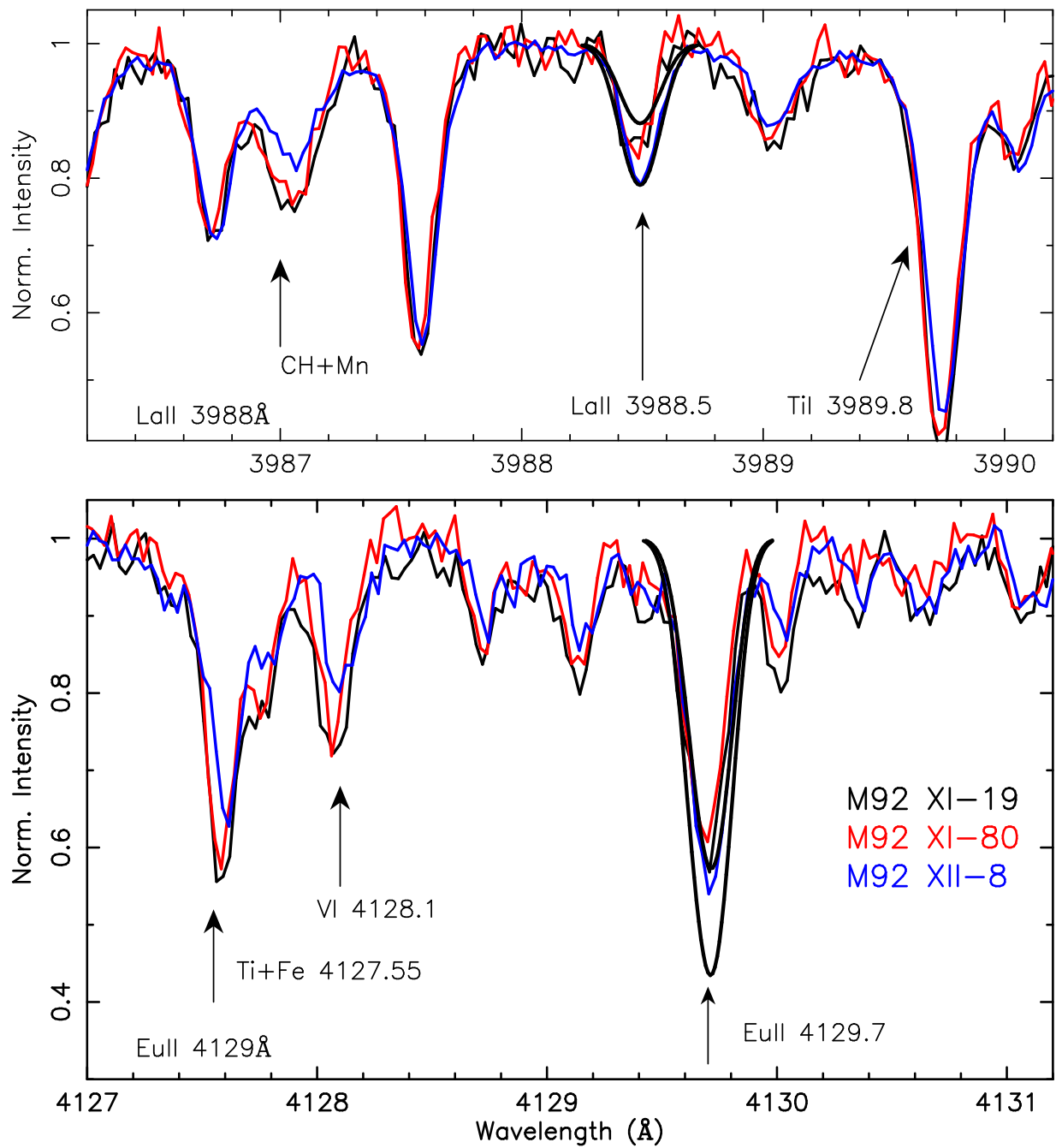}
\caption[]{Spectra for the three luminous red giants
M92  XI-19, XI-80, and XII-8 are superposed for the region of the
3988~\AA\ LaII line and  the
4129~\AA\ EuII line. These stars have $V$ between 12.78 and 13.09~mag. 
These three stars
were included in the sample of RS11; spectra of two
of them in the vicinity of the same La~II line and 
a Eu~II line at 3907~\AA\ are shown in their Fig.~7; 
they claim to have detected a very
large difference in rare earth abundances between the two stars.
The two thick black line profiles in each panel were synthesized using 
abundances of La or Eu differing by 0.3~dex.
\label{figure_m92_spec}}
\end{figure}

To illustrate the high S/N
of these spectra, sections around two key rare earth features
are shown in Fig.~\ref{figure_m92_spec}
for three of the sample red giants.
These three stars are in common with the sample of RS11, who
claim that they show a large range of rare earth abundances.
The two thick black curves  are spectral syntheses
for the two features with the abundance of the relevant element, La or Eu,
altered by 0.3~dex.

We selected the strongest and least crowded lines of
YII,  BaII, LaII, and EuII within the wavelength regime
covered.  We chose the 5854~\AA\ line of BaII
instead of the stronger line at 4554~\AA\ as it is 
very strong in the coolest stars.
Hyperfine structure patterns for BaII
are from \cite{mcwilliam98}.  
For Eu~II, we chose the resonance line at 4129~\AA\ as it is
both stronger and  less
crowded than the bluer Eu~II  lines, including that used by RS11 at 3907~\AA.  
The transition probability and HFS are
from \cite{europium}.  The Eu~II lines are
partially resolved due to the wide HFS pattern; the HFS correction
for the 4129~\AA\ line
ranges from $-1.1$~dex for the coolest M92 giant in our sample to zero
for the weak line seen in the hottest giants in our sample.
The wide HFS pattern for Eu explains why the change in line
strength 
shown in Fig.\ref{figure_m92_spec} 
for a line well above the linear part of the curve of growth
is so large for a 0.3~dex
change in Eu abundance.
We chose La~II 3988~\AA, 
one of the four lines used by RS11,
with a $gf$ value and HFS pattern from \cite{lanthanum}.  This line is
too weak to measure reliably in the two hottest stars in our sample
as well as in M92 IV-10. 
Among the cooler M92 giants, the HFS correction
for the 3988~\AA\ feature of La~II varies from $-0.17$~dex to
$-0.07$~dex.
The $gf$ values for the two Y~II
lines are taken from the current version of NIST \citep{nist}.
HFS corrections are not needed for Y~II \citep{hannaford}.

We measured equivalent widths for the same small set of rare
earth lines in each of the M92 giants.
The agreement of the measured $W_{\lambda}$ from the
two  spectra of M92 XII-8 taken 5 years apart is such that
we adopt an uncertainty in these measurements of 3~m\AA.  For
the two warmer stars which are more than 2~mag fainter in $V$
and have much weaker rare earth lines
we adopt 4~m\AA\
as the uncertainty in $W_{\lambda}$.  The region of the
4129~\AA\ line of Eu~II becomes
rather crowded in the two coolest stars and its $W_{\lambda}$
is somewhat more uncertain for them.  
We also measured equivalent widths
of any other lines of these four elements that we could detect.
HFS patterns from the sources
given above were used for all lines of BaII, LaII, and
EuII; $r$-process isotopic ratios were assumed.

The abundance analysis follows our previously published work,
see, e.g. \cite{cohen05a}.  We use a photometric definition
for $T_{eff}$ using $V-I$, $V-J$, and $V-K_s$, where the optical
colors are from \cite{stetson05} and the infrared ones from 2MASS
\citep{2mass1,2mass2}.  
We use the predicted color grid of
\cite{houdashelt00}.   We adopt the mean of the three resulting
values of $T_{eff}$ for our abundance analysis, then derive
a surface gravity assuming a mass along the RGB of 0.8~$M_{\odot}$,
with the distance and interstellar absorption to M92 taken from \cite{harris96}.
We adopt these computed values without modification
for all stars.
The resulting stellar parameters, and the dispersions of the three
values for $T_{eff}$, are given in Table~\ref{table_m92_sample}.
We use 1D LTE stellar model atmospheres from \cite{kurucz93}
with the abundance analysis code MOOG \citep{moog}.  The
microturbulent velocity $v_t$ is set by requiring the derived
abundance for individual Fe~I lines to be independent of
the equivalent width.   

This is a differential analysis from star to star within M92.
The uncertainties for the
absolute abundances of the rare earths for each star will have a term
representing the uncertainty of the relative $T_{eff}$,
a contribution from the uncertainty in the equivalent widths,
and a contribution from a possible small error in $v_t$.
We assign an uncertainty of 30~K to the $T_{eff}$ based
on the dispersion in this parameter as determined
from the various available colors,  given in Table~\ref{table_m92_sample}.
This translates into an  uncertainty in the absolute 
abundances for the rare
earth elements under consideration of a maximum of 0.04~dex, including
the contribution from the matching log($g$) change.
The contribution for an error in $v_t$ of 0.1 km s$^{-1}$ depends
on the line strength and whether or not HFS corrections
apply.  It appears to be largest for 5854~\AA\ Ba~II line (0.07~dex) and
drops to 0.02~dex for the other rare earth species under consideration.

For the two warmer stars, the dominant uncertainty arises from the
4~m\AA\ allowance for errors in the measured equivalent widths.
Summing the three terms in quadrature, for these two stars we adopt
an uncertainty in absolute abundance for the rare earth species studied
here of 0.13~dex.  For the cooler stars, the three terms contribute
roughly
equally, and we adopt an uncertainty in absolute abundances
of 0.08~dex for the species of interest.  We increase the overall
uncertainty in absolute abundances by 15\% for the three M92 giants
not included in the photometric database of \cite{stetson05}.

Table~\ref{table_m92_abund}
provides statistics for the absolute abundances
of these four rare earths in our sample of M92 luminous giants
for both only the best lines and for when all available
lines are used.
The maximum dispersion is only 0.07~dex for these 4 elements
within the 12 stars studied in either case.  The dispersion
for the Fe abundance as derived from either the neutral or
singly ionized species is also 0.07~dex.

\begin{figure}
\epsscale{1.0}
\plotone{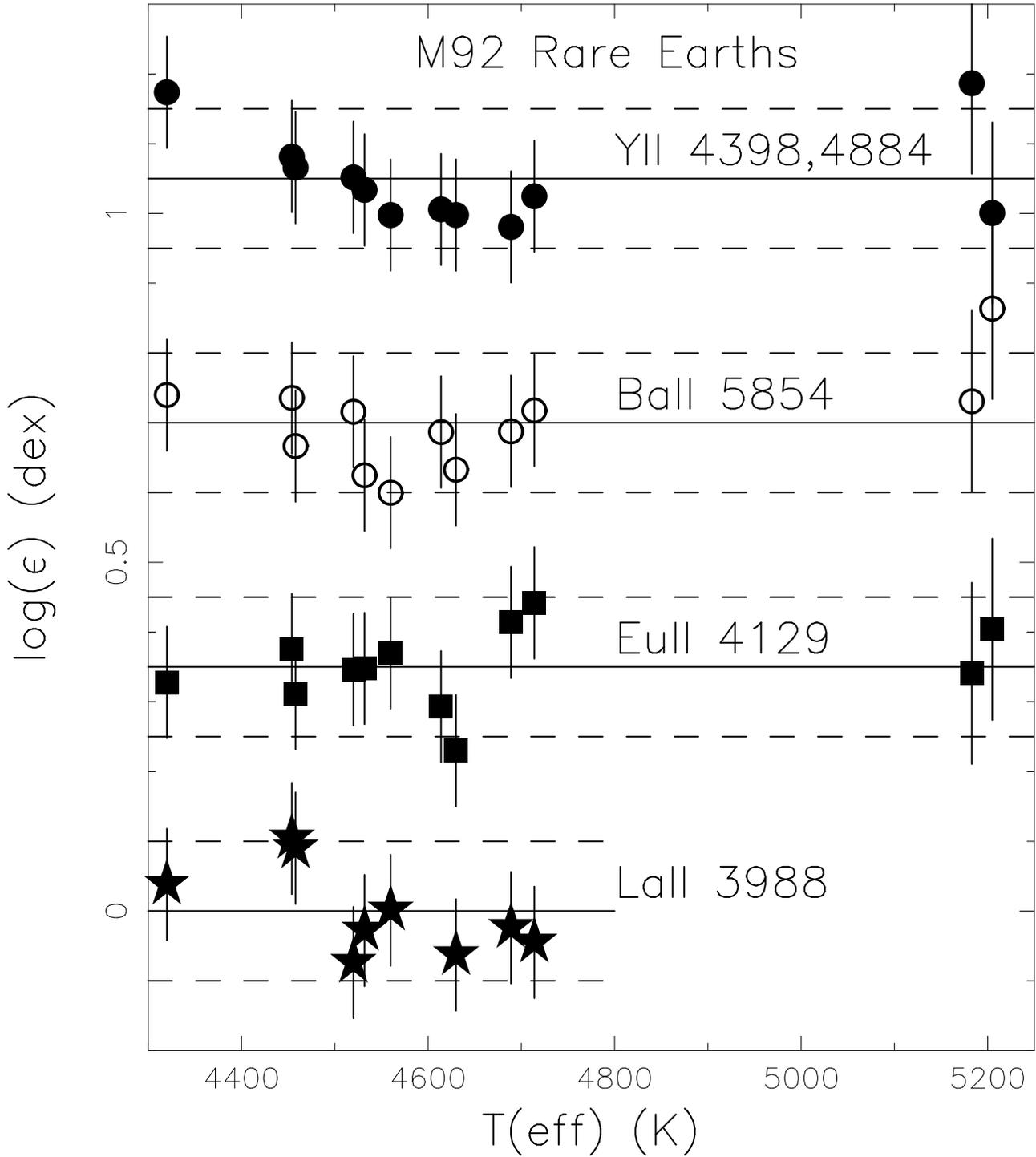}
\caption[]{Abundances deduced from the best rare earth lines
are shown for 12 luminous red giants in M92. The solid horizontal line
represents the mean abundance of the element, while the dashed lines
are offset above and below the mean by 0.1~dex.  The vertical axis values
are offset by a  constant between each line.
\label{figure_m92_abund}}
\end{figure}

\begin{figure}
\epsscale{1.0}
\plotone{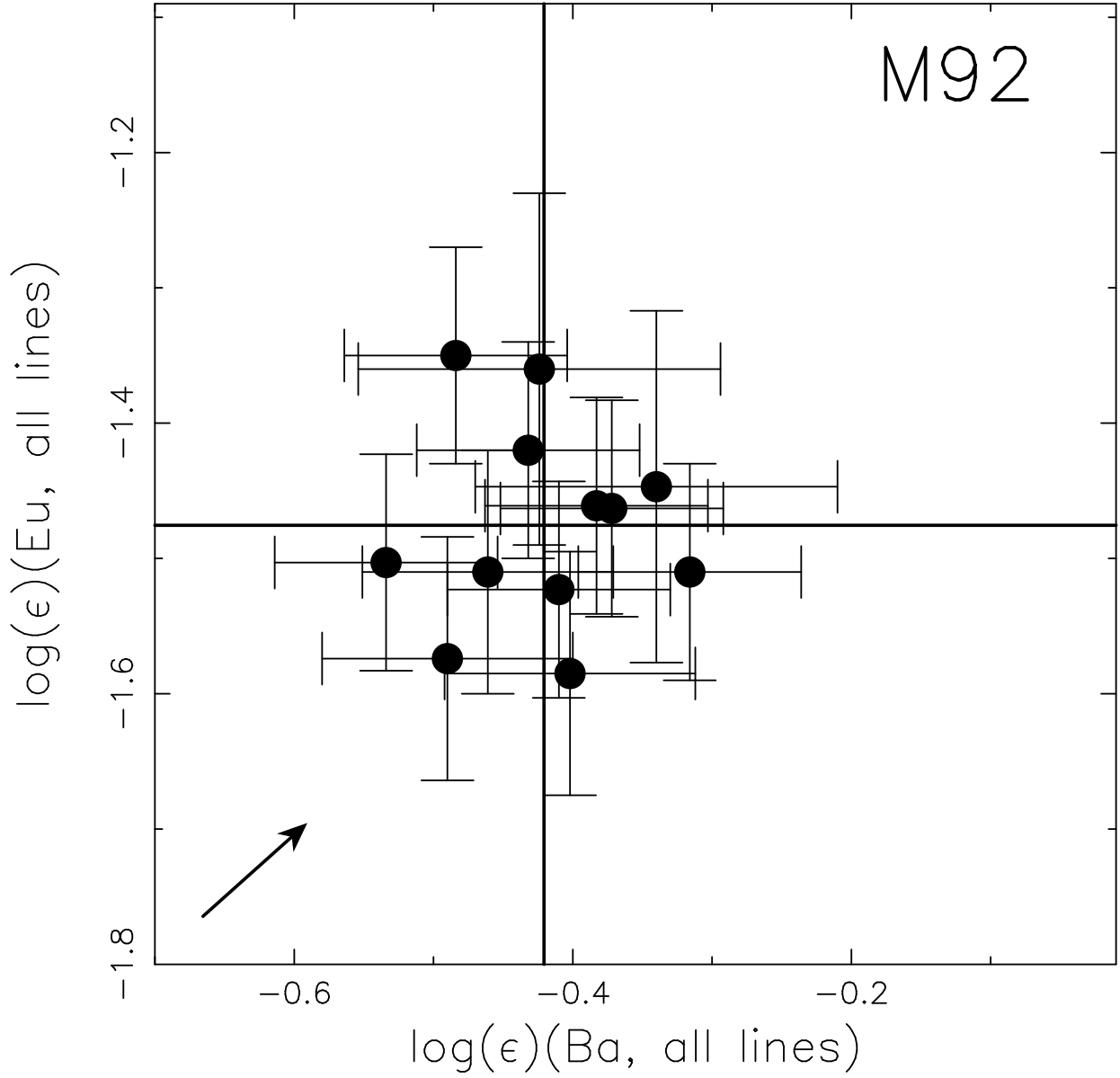}
\caption[]{Ba versus Eu abundances are  shown for our M92 sample.
The arrow at the lower left indicates the effect of a $T_{eff}$
increase of 50~K.
\label{figure_ba_eu}}
\end{figure}

The derived absolute abundances of Y, Ba, La, and Eu for each
of the 12 luminous red giants in M92 using the best
lines for each of these elements are shown in
Fig.~\ref{figure_m92_abund}.  Fig.~\ref{figure_ba_eu}
shows the derived absolute abundance of Eu versus that of Ba.
There is no sign of any correlation.
Furthermore a comparison of the synthesized
and observed line profiles
for three stars with very close $T_{eff}$
for the 4129~\AA\ Eu~II and
the 3988~\AA\ La~II lines (Fig.~\ref{figure_m92_spec}) again
suggests very strongly that the range of abundance in the
heavy neutron capture elements among the M92 sample giants
must be small.  

On the basis of this evidence we conclude that 
the abundances of each of
these four heavy neutron capture elements are constant
from star to star within our sample to within the expected
uncertainties discussed above.   The almost constant Eu abundance
derived from the 4129~\AA\ line is very gratifying given the very large
HFS correction for this strong line in the coolest sample stars, while it
is negligible in the faintest (hence warmest) two stars.  The
same comment applies for the four Ba~II lines used, as the line at 4554~\AA\
has a substantial HFS correction while the others do not.

It is interesting to compare the abundance ratios for these
four heavy neutron capture elements with those of
other Galactic globular clusters of similar metallicity.
J.~Cohen and collaborators \citep{m30} have recently carried out 
in an identical manner a detailed abundance
analysis for a sample of luminous
red giants from  another nearby
metal-poor  old Galactic GC,  M30.
Their [Fe/H] values are almost
the same, $-2.33$~dex for M92 vs $-2.40$~dex for M30.  Their
ratios $<$[Y/Fe]$>$ and $<$[Ba/Fe]$>$ are identical to within 0.04~dex.  That
for [La/Fe] is more discrepant, but the La~II lines are
weak in such metal-poor stars, and the resulting
La abundances uncertain unless the spectra are of very high S/N.
$<$[Eu/Fe]$>$ differs only by 0.09~dex.  For each of these four
heavy neutron capture elements the ratios agree to within the
uncertainties between the M92 sample presented here
and the  M30 sample of \cite{m30}.

The neutron capture element abundances in both M92 and in M30
are indicative of the $r$-process.  The high enhancement
of [Eu/Fe], about +0.5~dex above that of [Y, Ba, La/Fe],
signals that the $r$-process dominates in M92, as is the case in other
very metal-poor globular clusters \citep[see e.g.][]{gratton_araa}.

\section{Comparison with the work of Roederer \& Sneden 
\label{section_comp_rs} }

The sample of RS11 is somewhat
larger than ours, 19 vs 12 M92 red giants.  However,
only 16 of the 19 stars in their sample actually have measured Eu
abundances; all 19 have Y and La abundances.
There are no suitable Ba lines within
 the limited spectral coverage of the RS11 spectra.
Six of the 12 M92 red giants in our sample are in common
with that of RS11.  One of these, M92 IV-10, has
a very low S/N spectrum in RS11; no
abundances are tabulated in their paper for this star. 
For the other 5 stars in common,
we find a total range in log[$\epsilon$(Eu)] of 0.11~dex,
while they find a total range of 0.44~dex.  For La~II,
our range for these 5 stars is only 0.07~dex,
while their total range is 0.19~dex.  RS11 state
that the Y abundance, inferred from two Y~II lines,
within their M92 sample is homogeneous; our result is in agreement
 with theirs.

We thus have a major disagreement on the range of the
Eu and, to a lesser extent, of the La abundances within the
respective samples of M92 giants between us and the very
recent work of RS11.   We examine the data and the
analysis techniques used by these two groups in an attempt
to isolate the cause of this disagreement.
The major difference between the present work and that 
RS11 is the higher quality of our Keck spectra.    
Their
wavelength range was restricted to 3850 to 4050~\AA\ with
a spectral resolution of 14,000.
Spectra with high S/N and  high spectral resolution
are key in any effort to work on these elements with only a few
detectable lines, most of which are rather weak.

Our techniques for stellar parameter determination are similar
to those of RS11.
However, we rely on the highly accurate 
photometric database of \cite{stetson05} to determine
$T_{eff}$ while they rely on the earlier photographic
study by \cite{buo83}.
Three stars in our sample chosen for observation in the early
summer of 2011 to increase
overlap with the sample of RS11 are not in the Stetson
database.  For these we resorted to the photometry of
\cite{buo83}.
A check of a sample of stars in common between these
two photometric studies shows that
the old photographic photometry of \cite{buo83} is systematically
fainter in $V$ by $\sim$0.06~mag, with a dispersion
of 0.03~mag.  The mean difference in $V$ corresponds to a 
$T_{eff}$ change of 50~K
for $V-K$ (75~K for $V-J$).  Use of the older photometry will thus
yield slightly lower $T_{eff}$ values for a given star.  Furthermore for 
these three stars, identified in Table~\ref{table_m92_sample},
there is no $I$ photometry, suggesting the possibility
of larger uncertainties in $T_{eff}$ for them. The older
photometry includes $B$, but we ignore $B-V$ colors as they lack sufficient
sensitivity to $T_{eff}$ and are not suitable for our purpose. 

A detailed star-by-star comparison of the stellar parameters
between the present work and that of RS11 suggests that
there is a systematic difference in the $T_{eff}$ scale with
theirs being somewhat cooler by $\sim$60~K.
However, in both cases the goal is  
a differential analysis from star to star within M92, and thus
this small systematic difference is irrelevant for present purposes.
We have also checked our HFS patterns versus those of RS11
and found that they are essentially identical.

We must thus conclude that the higher quality of the Keck HIRES spectra
has enabled us to reach a level of accuracy which was not possible
with the lower dispersion and limited spectral range of the
RS11 sample.  They claim detection of a ``clear star-to-star
dispersion spanning 0.5-0.8~dex'' in La, Eu, and Ho within their M92 sample.  
We have demonstrated that this is not correct for either La or Eu.
Ho has even weaker and more blended lines, so 
their claimed high dispersion of Ho within M92 must
be regarded with suspicion.

\section{Discussion }

We have established, in contradiction to the results of
RS11, that there is no detectable range in star-to-star abundance
of the heavy neutron capture elements Y, Ba, La, and Eu
in M92 within our sample
of 12 red giant cluster members which spans a range
of 3.4~mag in $V$ and  $\sim$900~K 
in $T_{eff}$.  Our dispersions in absolute abundances for
these 4 elements are $\leq$0.07~dex, consistent with the expected
observational and analysis uncertainties.
This is very gratifying given that some of the key spectral
features used have very wide hyper-fine structure patterns and hence
HFS corrections which strongly vary within our sample of M92 giants.
These elements show a $r$-process abundance distribution, as expected.
The heavy neutron capture elements within M92 show
a chemical inventory very similar to that of M30,
a slightly more metal-poor and slightly less massive
Galactic globular cluster.

So the remaining question is why is M15 the only globular cluster
known to date
within which the $r$-process dominates and
a dispersion of heavy neutron capture element
abundances exists.  Beyond invoking the high luminosity
and mass of M15, there is as yet no obvious answer.  A search
for abundance dispersions within massive but less well studied
globular clusters is now underway.

\acknowledgements

We are grateful to the many people who have worked to make the Keck
Telescope and its instruments a reality and to operate and maintain
the Keck Observatory.  The author wishes to extend special thanks to
those of Hawaiian ancestry on whose sacred mountain we are privileged
to be guests.  Without their generous hospitality, none of the
observations presented herein would have been possible.   The author
thanks NSF grant AST-0908139 for partial support.

{}

\clearpage

\begin{deluxetable}{l c rr rrrr r r}
\tablewidth{0pt}
\tablecaption{
\large{M92 Sample for Rare Earth Study
\label{table_m92_sample} } }
\tablehead{
\colhead{ID} & \colhead{$V$\tablenotemark{a}} & 
  \colhead{$T_{eff}\tablenotemark{b}$} &
  \colhead{log($g$)} & \colhead{$v_t$} & [Fe12/H]\tablenotemark{c} \\
\colhead{} & \colhead{(mag)}  & \colhead{(K)} & \colhead{(dex)}
 & \colhead{(km s$^{-1}$)}
}
\startdata
X-49 &  12.31 & 4320 (25) & 0.68 & 2.1 & 5.22 \\
III-65 & 12.42 & 4454 (30) & 0.83 & 2.3 & 5.14  \\
II-53 & 12.45 & 4458 (33) & 0.84 & 2.3 & 5.16 \\
XII-8 & 12.78 & 4520 (25) & 1.02 & 2.1 & 5.04 \\ 
XI-80 &  12.82\tablenotemark{d} & 4532 (35) & 1.12 & 2.0  & 5.17 \\
XI-19 &  13.09\tablenotemark{d} & 4560 (35) & 1.07 & 2.0 & 5.23 \\
IV-10 &  13.46\tablenotemark{d} & 4614 (33) & 1.35 & 2.0 & 5.20 \\
III-82 & 13.30  & 4630 (30) & 1.31 & 2.0 & 5.10 \\
IV-79 &  13.45 & 4689 (37) & 1.39 & 1.9 & 5.12 \\
XII-34 & 14.33 & 4722 (58) & 1.40 & 2.0 & 5.14 \\
X-20 & 15.57  & 5183 (35) & 2.48 & 1.5  & 5.16 \\ 
VI-90  & 15.69 & 5215 (24) & 2.53 & 1.6 & 5.13\\
\enddata
\tablenotetext{a}{Photometry from \cite{stetson05}.}
\tablenotetext{b}{The dispersion around the mean $T_{eff}$ from the
three colors used is given in parentheses.}
\tablenotetext{c}{The average of [Fe/H] from neutral and from ionized Fe lines.}
\tablenotetext{d}{Photometry from \cite{buo83}.}
\end{deluxetable}

\clearpage

\begin{deluxetable}{l c rr rrrr }
\tablewidth{0pt}
\tablecaption{
\large{Statistics for Absolute Abundances of Rare Earths in M92
\label{table_m92_abund} } }
\tablehead{
\colhead{Species} &  \colhead{Lines} & \colhead{Num.} &
\colhead{Mean} &  \colhead{$\sigma$} & \colhead{Min log($\epsilon$)} &
\colhead{Max log($\epsilon$)} & \colhead{$<$[X/Fe]$>$\tablenotemark{a} } \\
\colhead{} &    \colhead{(\AA)} & \colhead{Stars} & \colhead{(dex)} &
\colhead{(dex)} & \colhead{(dex)} &  \colhead{(dex)} &  \colhead{(dex)} 
}
\startdata 
FeI &   $\sim$70 & 12 & 5.14 & 0.07 \\
FeII &  $\sim$10 & 12 & 5.16 & 0.07 \\
YII & 4398, 4884 & 12 & $-0.28$ & 0.07 & $-0.34$ & $-0.15$ & $-0.19$ \\
YII & All\tablenotemark{b} & 12 & 
     $-0.36$ & 0.07 & $-0.43$ & $-0.17$ & $-0.27$ \\
BaII & 5854 & 12 & $-0.45$ & 0.07 & $-0.55$ & $-0.29$ & $-0.25$ \\
BaII & All\tablenotemark{c} & 12 & 
    $-0.42$ & 0.06 & $-0.50$ & $-0.32$ & $-0.22$ \\
LaII & 3988 & 9 & $-1.36$ & 0.07 & $-1.43$ & $-1.25$ & $-0.21$ \\
LaII & All\tablenotemark{d}& 9 &
    $-1.33$ & 0.06 & $-1.44$ & $-1.27$ & $-0.18$ \\
EuII & 4129 & 12 & $-1.50$ & 0.06 & $-1.56$ & $-1.41$ & +0.32 \\
EuII & All\tablenotemark{e} & 12 & 
       $-1.48$ & 0.07 & $-1.59$ & $-1.35$ & +0.34 \\
\enddata
\tablenotetext{a}{Assumes [Fe/H](M92) = $-2.33$~dex.}
\tablenotetext{b}{Up to 8 YII lines}
\tablenotetext{c}{The two bluer or all four of 4554, 5854, 6141, and 6196~\AA\ lines.}
\tablenotetext{d}{Up to 7 LaII lines.}
\tablenotetext{e}{All detected among the three 
EuII lines 3905, 4129,
6645~\AA.}
\end{deluxetable}

\clearpage

\end{document}